\documentclass[11pt]{article}
\usepackage[utf8]{inputenc}

\usepackage{a4wide}

\usepackage{amsfonts}
\usepackage{bbm}
\usepackage{amsmath}
\usepackage{mathtools}
\usepackage{amsthm}

\usepackage{amssymb}
\usepackage{booktabs}
\usepackage{longtable}
\usepackage{enumerate}
\usepackage{dsfont}

\usepackage{mathrsfs}

\usepackage[short,nocomma]{optidef}
\usepackage[ruled,noend,lined]{algorithm2e}
\SetKwComment{Comment}{$\triangleright$\ }{}
\SetKwInOut{KwIn}{Input}
\SetKwInOut{KwOut}{Output}

\usepackage{hyperref}
\usepackage{cleveref}

\newcommand{\E}{{\rm \mathbb{E}}}
\newcommand\dif{\mathop{}\!\mathrm{d}}
\newcommand*{\ratio}{4.39}
\newcommand*{\pars}{N}
\newcommand*{\card}{\mathcal{K}}
\newcommand*{\ctresh}{0.47521}
\newcommand*{\dtresh}{0.60138}
\newcommand*{\mubar}{0.72428}
\newcommand*{\ctreshfull}{0.4752190514489393}
\newcommand*{\dtreshfull}{0.6013835675554252}
\newcommand*{\fullratio}{4.383238341343964}

\newcommand*{\dminshort}{0.20319}
\newcommand*{\firstcondition}{0.02949}
\newcommand*{\secondcondition}{4\cdot10^{-6}}
\newcommand{\integralratio}{6.65}
\newcommand{\previousratio}{9.37}
\providecommand{\variable}{x}
\providecommand{\solfk}{\variable^*}
\providecommand{\utilization}{\mu}
\providecommand{\fractionofcapacityweightratio}{\delta}
\providecommand{\spaceleftmultiplier}{\frac{1}{1-\fractionofcapacityweightratio}}

\newcommand{\intInterCurly}[2]{\ensuremath{\{#1,\ldots,#2\}}}
\providecommand{\algsecretary}{\mathcal{A_S}}
\providecommand{\algknapsack}{\mathcal{A}_K}
\newcommand{\permprofitdef}{\sigma}
\newcommand{\permprofit}[1]{\permprofitdef(#1)}

\newcommand{\opt}{\operatorname{OPT}}

\newtheorem{theorem}{Theorem}

\newtheorem{lemma}[theorem]{Lemma}

\newtheorem{definition}[theorem]{Definition}

\newtheorem{fact}[theorem]{Fact}

\author{
Jeff Giliberti \\
ETH Zurich, MPI for Informatics\thanks{The work was carried out during a virtual internship at MPI for Informatics.} \\
\texttt{jgiliberti@student.ethz.ch}
\and
Andreas Karrenbauer \\
MPI for Informatics \\
\texttt{karrenba@mpi-inf.mpg.de}
}

\date{}

\title{Improved Online Algorithm for Fractional Knapsack in the Random Order Model\texorpdfstring{\thanks{This paper has been accepted at WAOA 2021.
The final authenticated publication is available online at \url{https://doi.org/10.1007/978-3-030-92702-8_12}}}{}}

\begin{document}
\maketitle

\begin{abstract}
    The fractional knapsack problem is one of the classical problems in combinatorial optimization, which is well understood in the offline setting. However, the corresponding online setting has been handled only briefly in the theoretical computer science literature so far, although it appears in several applications. 
    Even the previously best known guarantee for the competitive ratio was worse than the best known for the integral problem in the popular random order model. 
    We show that there is an algorithm for the online fractional knapsack problem that admits a competitive ratio of $\ratio$. 
    Our result significantly improves over the previously best known competitive ratio of $\previousratio$ and surpasses the current best $\integralratio$\nobreakdash-competitive algorithm for the integral case.
    Moreover, our algorithm is deterministic in contrast to the randomized algorithms achieving the results mentioned above.
\end{abstract}

\section{Introduction}

The knapsack problem is well-studied and has a long history in the literature, both the offline and the online versions, where in the latter the items are revealed one after the other and an irrevocable decision whether to pick the current item has to be made immediately. 
In the online setting, one typically considers the \emph{random order model}, in which the adversary controls the instance, i.e., the items with their weights and profits, but a permutation, chosen uniformly at random, determines the arrival order of the items.
In this model, the performance measure of an algorithm $\mathcal{A}$ is its \emph{competitive ratio}. We say that $\mathcal{A}$ is $r$\nobreakdash-competitive with respect to the profit $\opt(\mathcal{I})$ of an optimal offline algorithm for input $\mathcal{I}$, if $\E[\mathcal{A}(\mathcal{I})] \ge (1/r - o(1)) \cdot \opt(\mathcal{I})$ holds for all inputs $\mathcal{I}$. Here, the expectation is taken over the random permutation as well as over random choices of the algorithm. The $o(1)$-term vanishes asymptotically with respect to the number of items in the input instance.
In a series of papers \cite{Albers2021,babaioff2007knapsack,kesselheim2014primal}, the competitive ratio for online knapsack has been improved to $\integralratio$, achieved by a randomized algorithm in~\cite{Albers2021}.
The online version of the fractional knapsack problem has not received much attention in literature as of yet.
An application of the online fractional knapsack problem was introduced in~\cite{rao}, where the items are articles that are presented in a newsfeed to busy readers. The scarce resource is time and the profit is the information gain. Upon arrival of an article, readers see its length and a cue about the content. They can discard the article right away based on this information, or start reading it. They can read it completely or discard it at any time. The information gain is assumed to be proportional to the reading time and the cue about its content. Discarded articles are not considered again. Moreover,~\cite{rao} contains a framework that turns any $\alpha$-competitive algorithm for the online knapsack problem into an $(\alpha+e)$-competitive algorithm for the online fractional knapsack problem.
The approach is based on flipping a biased coin (w.r.t.\ $\alpha$) and either executing the given algorithm or the famous secretary algorithm, where the first $n/e$ items are discarded and then the first item that is better than the best item seen so far is chosen.
Together with the result of~\cite{Albers2021}, this led to the previously best known competitive ratio of $\previousratio$ for the online fractional knapsack problem. In this paper, we cut the competitive ratio by more than half, namely to $\ratio$.
To this end, we observed that the algorithm in~\cite{Albers2021} already contains a secretary algorithm on its own and that small and large items are treated separately due to the nature of the integral knapsack solution. Instead of treating it as a black box, relaxing the integrality constraints allows us to unify the small-item and large-item cases and output fractional values instead of binary ones, thereby achieving a better competitive ratio. 
Moreover, this also removes the randomization from the algorithm making it  deterministic, and the expected value that determines the competitive ratio solely depends on the randomness of the input permutation.

\subsection{Related Work}
The study of the online (integral) knapsack problem was initiated by Marchetti-Spaccamela and Vercellis~\cite{Marchetti-Spaccamela1995}, who showed that there is no constant-competitive deterministic online algorithm. Moreover, Chakrabarty et al.~\cite{ChakrabartyKnapsackRange} extended the same hardness result to randomized algorithms. Motivated by the difficulty of the adversarial model for this problem, a number of beyond worst-case scenarios have been investigated. The most popular of these is the random order model that has received increasing attention in the field of online algorithms. In this model, the online knapsack problem was first studied by~\cite{babaioff2007knapsack}, showing a $10e$-competitive algorithm. Kesselheim et al.~\cite{kesselheim2014primal} developed an $8.06$-competitive randomized algorithm for the generalized assignment problem, which generalizes to a setting with multiple knapsacks of different capacities. Albers et al.~\cite{Albers2021} achieved the currently best known upper bound of $6.65$.
The online fractional variant under adversarial arrivals was first considered in~\cite{ofkpr}. There, the knapsack capacity is augmented by a factor $1 \le R \le 2$, and the algorithm can remove previously packed items. The plain version of the fractional knapsack problem in the random order model was studied in~\cite{rao}, presenting a $\previousratio$-competitive randomized algorithm, which we improve upon. 
Recently,~\cite{10.1145/3428336}~considered a general version of the fractional online knapsack problem with multiple knapsacks and rate constraints.
A further branch of research considers the infinitesimal assumption (or that the packing is allowed to be \textit{fractional}), i.e., the profit of a single item is small compared to the profit of the optimal integral solution, under which Vaze~\cite{Vaze2016KnapsackMatching} gave a $2e$-competitive algorithm. In addition, a common approach combined with the infinitesimal assumption is to assume that the density (profit-size ratio) of each item is in a known range~\cite{buchbinder2005online,buchbinder2006improved,ChakrabartyKnapsackRange}. 
Another problem closely related to the random order model is the secretary problem~\cite{Dynkin63,lindley1961dynamic}, that is, a special case of the online knapsack problem in which the weights are uniform and equal to the weight constraint. One natural generalization of the latter is the $k$-secretary problem~\cite{ALBERS2021102,Chan2015strong2k,kleinberg2005Sum}, where $k$ elements need to be selected, as well as the matroid secretary problem~\cite{babaioff2018matroid,feldman2014simple}, where elements of a weighted matroid arrive in random order, and in both of these, the goal is to maximize the combined value of the selected elements. 
Other variants of online knapsack presented in the literature include removable models, where removals can incur no cost or a cancellation cost \cite{babaioff2008selling,babaioff2009selling,han2014online,han2015randomized,10.1007/3-540-45465-9_26}, reservation costs~\cite{bockenhauer_et_al:LIPIcs.STACS.2021.16}, an expected capacity constraint~\cite{vaze2018online}, and resource buffering~\cite{han_et_al:LIPIcs:2019:11524}.
Furthermore, there have been alternative approaches to the random order model, such as \textit{stochastic} versions of the online knapsack problem~\cite{dean2005adaptivity,doi:10.1287/moor.1080.0330,GOERIGK201512,LUEKER1998277,marchetti1995stochastic}, a random order model with bursts of adversarial time steps~\cite{kesselheim_et_al:LIPIcs:2020:12479}, and the
the advice complexity model~\cite{BOCKENHAUER201461}.

\subsection{Our Contribution}
\begin{theorem}\label{thm:main}
    There exists a $\ratio$-competitive deterministic algorithm for the online fractional knapsack problem in the random order model.
\end{theorem}
We achieve this result by analyzing a natural variation of the algorithm from~\cite{Albers2021}. As in the original version for the online knapsack problem, the algorithm works in three phases -- the sampling phase, the secretary phase, and the knapsack phase. The transition between the phases happens at iterations $\lfloor cn\rfloor$ and $\lfloor dn \rfloor$, respectively, where $0 < c \le d \le 1$ are optimized to achieve the best possible competitive ratio when combining our analyses of the second and third phase contributions. The algorithm can be considered as a blending of two algorithms that share the first phase, which is possible because no items are picked in the sampling phase; thus, there is no interference between both algorithms. After the first phase, i.e., after roughly 47.5\% of all items, the secretary algorithm takes over and exclusively decides the items of the second phase. It does so by selecting all items that have a larger profit than the most profitable item seen in the sampling phase. Should picking the current item exceed the knapsack capacity, the item is picked to the largest extent possible to fill the knapsack. If there is still capacity left after the second phase, i.e., after roughly 60.1\% of all items, we switch from the secretary to the knapsack algorithm, which then exclusively fills in a fraction of each arriving item according to the optimal (fractional) solution of all items revealed so far and the remaining capacity. It is interesting to note that we do not distinguish between large and small items in our analysis, in comparison to~\cite{Albers2021}. Furthermore, in the analysis of the secretary algorithm, we only account for the probability of picking the item that has the largest contribution to the objective value of the optimal (fractional) solution. This is sufficient to cover the case where the optimal solution consists of a single item, a situation that the knapsack algorithm (or our analysis of it) cannot handle well. In fact, if we forced $c=d$ in the parameter optimization, we could not get a better competitive ratio than $6.63$, which is still better than the competitive ratio from~\cite{rao}, but significantly worse than the best ratio from this paper. On the other hand, if we did not use the knapsack algorithm at all ($d=1$), we obtained the well-known secretary algorithm, where the sampling phase ends after skipping a $c=1/e$ fraction of all items (justifying the naming). 

\section{Preliminaries}
\begin{definition}[\textbf{OFKP}] 
    We are given a set $I$ of $n$ items, each item $i \in I$ has \textit{size} $s_i \in \mathbb{Q}_{>0}$ and a \textit{profit} (\textit{value}) $v_i \in \mathbb{Q}_{\ge 0}$. The goal is to find a maximum profit fractional packing into a knapsack of size $W\in \mathbb{Q}_{>0}$, i.e., a solution $\variable \in \mathbb{Q}_{\ge 0}^n$ s.t. $\sum_{i\in I} s_i\variable_i \le W$ and $\sum_{i\in I} v_i\variable_i$ is maximized. The items are revealed one by one in a round-wise fashion. In round $\ell \in [n]$, the algorithm sees item $\ell$ with its size and profit. It has to decide immediately and irrevocably the fraction $\variable_\ell$ of the current item in the final packing.
\end{definition}
We make the following two assumptions without loss of generality: (i) No item has size larger than the knapsack capacity\footnote{Items whose size exceeds the capacity of the knapsack can be cut at the knapsack capacity.}, (ii) Items have distinct values\footnote{It can be accomplished in polynomial time by fixing a random (but consistent) tie-breaking between elements of the same value, based for instance on the identifier of the element~\cite{babaioff2007knapsack}.}. Using assumption (ii), we obtain that there is a unique (optimum) solution $\solfk$ for the given set of items $I$. In fact, profit-to-weight ratio ties can be broken by taking the most valuable element which is unique by assumption (ii). 

Next, we formalize the optimal (fractional) solution for a given subset $Q \subseteq I$. Let the \textit{density} of an item be the ratio of its profit to its size. The optimal (fractional) solution has a clear structure: There exists a density threshold $\rho_Q$ such that any item $i \in Q$ with $v_i / s_i > \rho_Q$ has $\variable_i = 1$ and any item $i \in Q$ with $v_i / s_i < \rho_Q$ has $\variable_i = 0$. Meaning that the $k-1$ densest items are packed integrally and the remaining space is filled by the maximum feasible fraction of the $k$-th densest item.
Let $I(\ell)$ denote the subset of items $I$ revealed up to round $\ell$ and define $\variable^{(\ell)}$ to be the optimal (fractional) solution for the item set $I(\ell)$. Let $\opt$ be the total profit of the optimal (fractional) solution $\solfk$ for the item set $I$, i.e., $\opt = \sum_{i\in I} v_i \solfk_i$. For convenience of notation, let $\opt$ also denote the set of items that are part of the optimal (fractional) solution, i.e., each item $i$ whose $\solfk_i > 0$, and let $\card$ be its cardinality. Additionally, we label items in descending order of contribution to $\opt$ such that $v_i \solfk_i \ge v_{i+1} \solfk_{i+1}$ for all $i \in [n-1]$. 

As in~\cite{Albers2021}, we will use the following well-known fact to obtain lower or upper bounds on sums in closed form.

\begin{fact} \label{fact:integralbounds}
 Let $f$ be a non-negative real-valued function and let $a,b \in \mathbb{Z}$.
    \begin{itemize}
        \item[(A)] If $f$ is non-increasing, then $\int_a^{b+1} f(t) \dif{t} \le \sum_{t=a}^b f(t) \le \int_{a-1}^b f(t) \dif{t}$.
        \item[(B)] If $f$ is non-decreasing, then $\int_{a-1}^{b} f(t) \dif{t} \le \sum_{t=a}^b f(t) \le \int_{a}^{b+1} f(t) \dif{t}$.
    \end{itemize}
\end{fact}

\subsection{Review of the Blended Approach}
A standard approach used by packing algorithms in the online setting is the following. Algorithms have a sampling phase, during which all items are rejected, and a decision phase, where items may be accepted according to some decision rule developed according to the information gathered in the sampling phase. The novel idea presented in~\cite{Albers2021} is to combine two algorithms, $\mathcal{A}_1$ and $\mathcal{A}_2$, in a blended manner. The strategy is to make a better use of the entire instance by letting the two algorithms have a common sampling phase, and, subsequently, using the sampling phase of one algorithm as the decision phase of the other. Let $0 < c \le d \le 1$ denote some constant parameters to be specified later. Rounds $1,\ldots,\lfloor cn \rfloor$ define the common sampling phase. For rounds $\lfloor cn \rfloor +1,\ldots, \lfloor dn \rfloor$, there is the $\mathcal{A}_1$ decision phase, while, $\mathcal{A}_2$ continues its sampling phase. From round $\lfloor dn \rfloor +1$ the algorithm $\mathcal{A}_1$ stops and the $\mathcal{A}_2$ decision phase starts until the end of the stream. As in~\cite{Albers2021}, we make the  assumption $cn, dn \in \mathbb{N}$ for the analysis, which does not affect the competitive ratio substantially for $n$ large enough. Clearly, combining the two algorithms without an initial random choice on whether to run $\mathcal{A}_1$ or $\mathcal{A}_2$ (\cite{babaioff2007knapsack,rao,kesselheim2014primal}) comes at the cost of possibly having a non-empty knapsack when $\mathcal{A}_2$ starts its decision phase, thus, deteriorating $\mathcal{A}_2$ performance. However, we will see that with some sufficiently high probability the algorithm $\mathcal{A}_1$ does not pack any item.

In the online integral knapsack algorithm developed in~\cite{Albers2021}, the algorithm $\mathcal{A}_\mathscr{L}$ deals with all items that consume more than $1/3$ of the knapsack capacity. Items whose size is at most $1/3$ of the knapsack capacity are packed by algorithm $\mathcal{A}_\mathscr{S}$. The algorithm $\mathcal{A}_\mathscr{L}$ exploits the connection with the $2$-secretary problem because at most two large items can fit in the knapsack. The algorithm $\mathcal{A}_\mathscr{S}$ packs the current small item integrally with probability equal to its value in the \textit{optimal fractional solution} formed by the items occurred so far. One possible reason for which $\mathcal{A}_\mathscr{L}$ and $\mathcal{A}_\mathscr{S}$ operate on different instances, i.e., the instances that consist of large and small items respectively, is that the optimal integral solution is not \textit{monotone}. Namely, in the optimal integral solutions for the items seen until rounds $\ell,m$ for $\ell < m$ denoted by $\variable^{(\ell)}$ and $\variable^{(m)}$ respectively, there might be an item $i$ such that $\variable_i^{(m)} = 1 > \variable_i^{(\ell)} = 0$. 
Notably, this comes at the cost of not packing large items using $\mathcal{A}_\mathscr{S}$, as the large items in the optimal integral solution that algorithm $\mathcal{A}_\mathscr{L}$ considers may be arbitrarily different from the large items in the optimal fractional solution that $\mathcal{A}_\mathscr{S}$ considers.

\subsection{Application of the Blended Approach in the Fractional Case}

In the fractional variant of the problem, we can overcome the limitation described above. The fractional relaxation allows us to make both algorithms sharing the same instance of the problem.
The \textit{secretary} algorithm $\algsecretary$ packs the most profitable items, whereas, the \textit{knapsack} algorithm $\algknapsack$ packs both small and large items (fractionally when needed) that are part of the optimal fractional solution $\opt$.
In the analysis of $\algsecretary$, the connection between the fractional knapsack problem and the $k$-secretary problem can be extended beyond the $2$-secretary problem, unlike the integral case, losing only the symmetry property of packing an item before another when there is an item taken fractionally in the optimal (fractional) solution. However, we will see that the connection with the $1$-secretary problem is enough for our purpose. Next, we give give a high-level description of the proof of Theorem~\ref{thm:main} using Algorithm~\ref{algo:main}.
\begin{algorithm}[ht] \label{algo:main}
\caption{Online Fractional Knapsack $\mathcal{A}$}
    \KwIn{permutation $\pi$ of item set $I$, knapsack capacity $W$, parameters $c,d \in (0,1]$ with $c \le d$.\\ Item $i$ appears in round $\pi(i)$ and reveals $v_i$ and $s_i$.}
    \KwOut{feasible fractional knapsack packing, i.e., $0 \le \variable_i \le 1$ for item $i \in [n]$.\medskip}
    Let $\ell$ be the current round and $i$ be the online item of round $\ell$\;\vskip 3pt
    Let $v^*$ be the maximum profit seen up to round $\lfloor cn \rfloor$\;\vskip 3pt
    \For(\Comment*[f]{Algorithm $\algsecretary$}){round $\ell \in \left\{\lfloor cn \rfloor +1, \ldots, \lfloor dn \rfloor\right\}$}{
        \uIf{$v_i > v^*$}{
            Set $\variable_i = \displaystyle\frac{1}{s_i} \cdot \min\left\{s_i,\: W-\displaystyle\sum_{j : \pi(j) < \ell}  s_j \variable_j \right\}$\;
        }
        \uElse{
            Set $\variable_i = 0$\;
        }
    }
    \vskip 3pt
    \For(\Comment*[f]{Algorithm $\algknapsack$}){round $\ell \in \left\{\lfloor dn \rfloor +1,\ldots,n\right\}$}{
        $\variable^{(\ell)} :=$ optimal fractional knapsack packing on $I(\ell)$\;
        Set $\variable_i = \displaystyle\frac{1}{s_i} \cdot \min\left\{s_i \variable^{(\ell)}_i,\: W-\displaystyle\sum_{j : \pi(j) < \ell} s_j \variable_j \right\}$\;
    }
\end{algorithm}
\begin{proof}
    Let $\mathcal{A}$ be the algorithm obtained by combining $\algsecretary$ and $\algknapsack$. Moreover, define $p_i, q_i$ to be some probabilities that we introduce later. By setting parameters $c,d$ to $\ctresh$, $\dtresh$, we will show
    \begin{align*} 
        \E[\mathcal{A}] \ge \E[\algsecretary] + \frac c d\cdot \E[\algknapsack] 
        &\ge \sum_{\substack{i\in \opt}} p_{i} \cdot  v_i \solfk_i  \:+\: \frac c d \sum_{i \in \opt} q_i \cdot  v_i \solfk_i \\
        &\ge \left(\frac{1}{\ratio} - o(1)\right)\cdot \opt.\qedhere
    \end{align*}
\end{proof} 
\pagebreak[2]
In the remainder of this paper, we analyze the performance of $\algsecretary$ and $\algknapsack$. The algorithm $\algsecretary$ and its analysis are similar to the algorithm $\mathcal{A}_\mathscr{L}$ and its analysis in~\cite{Albers2021} based on the \textsc{single-ref} algorithm~\cite{ALBERS2021102} for the $k$-secretary problem. The algorithm $\algknapsack$ and its analysis extend the approach of~\cite{Albers2021} to consider the possibility of packing items fractionally. Namely, we make $\algknapsack$ taking the largest possible fraction of an item that is part of the optimal fractional solution seen at round $\ell$ according to the available knapsack capacity. The resulting competitive ratio obtained by combining lower bounds on the expected profit of algorithms $\algsecretary$ and $\algknapsack$ is analyzed in Section~\ref{sec:cra}.

\section{Secretary Algorithm $\algsecretary$}
The following is an adaptation of \textsc{single-ref} developed for the $k$-secretary problem in~\cite{ALBERS2021102} and applied to the knapsack setting in~\cite{Albers2021}. There is a useful connection between the online knapsack problem under random arrival order and the $k$-secretary problem. The latter is defined as an unweighted version of online knapsack in the random order model, or equivalently, items can be seen as $W/k$ large. In contrast, in our problem items may be larger than $W/k$. 
\paragraph{Algorithm.}
The algorithm $\algsecretary$ works as follows. During the initial sampling phase of $\lfloor cn \rfloor$ rounds, the algorithm rejects all items and identifies as \textit{best sample} the encountered element with the highest value. In rounds $\ell \in \{\lfloor cn \rfloor + 1, \ldots, \lfloor dn \rfloor\}$, the algorithm takes the the largest possible fraction (according to the remaining knapsack capacity) of the items whose individual profit beats the profit of the best sample. Consequently, the first item beating the best sample will be taken integrally. Then, the algorithm maximizes the fraction of each subsequently accepted item.

\paragraph{Analysis.}
Let $p_i$ for $i \in [n]$ be the probability that $\algsecretary$ packs the $i$-th most profitable item as the first element. When considering the probability $p_i$, we do not specify what the algorithm will further do, i.e., after the first accepted item, any or no second item may be included and so on for subsequent items. 
In the following, we report the lower bounds for the probabilities $p_i$ showed in~\cite{Albers2021}. Note that these lower bounds do not depend on the integral or fractional variant considered as the \textit{first} item can be always packed integrally. Let us define a permutation $\permprofitdef : [n] \rightarrow [n]$ such that $v_{\permprofit{1}} > v_{\permprofit{2}} > \ldots > v_{\permprofit{n}}$.
\begin{lemma}[\cite{Albers2021}]\label{lemma:largeprob} We have the following lower bounds for the probability $p_i$ that item $\permprofit{i},\, i \in [n]$, is accepted by $\algsecretary$ as the first item
    \begin{align*}
        p_i \ge p(i) - o(1), \text{ with }p(i) = c\ln \frac d c + c \sum_{k=1}^{i-1} \binom{i-1}{k} (-1)^k \frac{d^k - c^k}{k}.
    \end{align*}
\end{lemma}
By the lemma above, we have $p(1) = c \ln \frac{d}{c}$ for $i=1$. Furthermore, the value of $\opt$ is upper bounded by the sum of the profits of the $\card$ most profitable items because items are sorted in decreasing order of contribution to $\opt$, thus, $v_{\permprofit{i}} \ge v_i \solfk_i$. This yields
\begin{align*}
    &\E[\algsecretary] \ge \sum_{i=1}^\card p_i v_{\permprofit{i}} \ge \sum_{i=1}^\card p_i  v_{i} \solfk_{i},
\end{align*}
concluding the analysis of algorithm $\algsecretary$.
\section{Knapsack Algorithm $\algknapsack$}
In this section, we present the algorithm $\algknapsack$ that leverages the structure of optimal fractional solutions restricted to the items seen so far. In contrast to the small-item algorithm developed in~\cite{Albers2021}, which uses the optimal fractional solutions to obtain an integral packing via randomized rounding, our deterministic algorithm uses the value of the current item in the fractional solution to the extent of the remaining knapsack capacity.
\paragraph{Algorithm.}
The algorithm $\algknapsack$ works as follows. During the initial sampling phase of $\lfloor dn \rfloor$ rounds, the algorithm rejects all items. In each round $\ell \ge \lfloor dn \rfloor + 1$, the algorithm computes an optimal fractional solution $\variable^{(\ell)}$ for $I(\ell)$. We pack an $\variable_i^{(\ell)}$ fraction of the current item if there is enough space, otherwise we pick the largest possible fraction according to the remaining space. Thus, the algorithm determines the fraction of item $i$, denoted by $\variable_i$, as follows 
\[
\variable_i = \frac{1}{s_i} \cdot \min\left\{s_i \variable_i^{(\ell)},\: W - \sum_{i \in I(\ell - 1)} s_j \variable_j \right\}.
\]
\paragraph{Analysis.}
We study the performance of algorithm $\algknapsack$ assuming it has the entire knapsack at its disposal, i.e., a capacity of $W$, and afterwards show how this occurs with constant probability. In our proofs, we consider an arbitrary fixed element $i \in \opt$ and define $\fractionofcapacityweightratio \in (0,1]$ to be a parameter representing the fraction of the knapsack capacity $W$ that element $i$ occupies in our online solution. 
For a fixed $\fractionofcapacityweightratio$, the proofs of Lemma~\ref{lemma:small-singleround} and Lemma~\ref{lemma:small-allrounds} almost immediately follow from the small-item case analysis in~\cite{Albers2021}; we reproduce their proofs for completeness and include few changes in Lemma~\ref{lemma:small-singleround} that have to be made in order to adapt them to our fractional setting. 
In the second part of the analysis, in Lemma~\ref{lemma:small-packing}, we make use of $\fractionofcapacityweightratio$ to exploit the possibility of packing items fractionally.
\begin{lemma}[\cite{Albers2021}]\label{lemma:small-singleround}
    Let $i \in \opt$ and $\variable_i(\ell)$ be the fraction of item $i$ that is packed by $\algknapsack$ in round $\ell$. For $\ell \ge dn + 1$, it holds that \[\Pr\left[\variable_i(\ell) \ge \min\left\{\frac{\fractionofcapacityweightratio W}{s_i},\, \solfk_i\right\} \right] \ge \frac 1 n \left(1 - \spaceleftmultiplier \ln \frac{\ell}{dn}\right).\]
\end{lemma}
\begin{proof}
    In a random permutation item $i$ arrives in round $\ell$ with probability $1/n$. 
    In round $\ell \ge dn + 1$, the algorithm packs $i$ for an $\variable_i^{(\ell)}$ fraction provided that there is enough space. Note that the rank w.r.t. profit-to-weight ratio of item $i$ in $I(\ell)$ is less than or equal to its rank in $I$. According to the structure of the optimal fractional solutions, this implies that $\variable_i^{(\ell)} \ge \solfk_i$. Moreover, if the current resource consumption $X$ is at most $(1-\fractionofcapacityweightratio)W$, then the current item $i$ can be packed up to a fraction $\delta W / s_i$. 
    By treating $\solfk_i$ as a parameter, it is not required to analyze the resource consumption in each round in expectation over all items. The latter approach appears in~\cite{KRTVowbm}, which relies on the fact that in any step $k$ of the algorithm the choice of the random permutation up to this point can be modeled as a sequence of independent random experiments.
    Let $X_k$ be the resource consumption in round $k < \ell$. By assumption, the knapsack is empty after round $dn$, thus $X = \sum_{k=dn+1}^{\ell-1} X_k$. Let $Q$ be the set of $k$ visible items in round $k$. The set $Q$ can be seen as uniformly drawn from all $k$-item subsets and any item $j \in Q$ is the current item of round $k$ with probability $1/k$.
    The algorithm packs any item $j$ for at most an $\variable_j^{(k)}$ fraction, thus
    \begin{equation*}
        \E[X_k] \le \sum_{j\in Q}\Pr[\text{$j$ occurs in round $k$}] s_j \variable_j^{(k)} = \frac{1}{k} \sum_{j\in Q} s_j \variable_j^{(k)} \le \frac{W}{k},
    \end{equation*}
    where the last inequality holds because $\variable^{(k)}$ is a feasible solution for the knapsack of size $W$. By linearity of expectation and the previous inequality, the expected resource consumption up to round $\ell$ is 
    \begin{equation*}
        \E[X] = \sum_{k = dn+1}^{\ell-1} \E[X_k] \le \sum_{k = dn+1}^{\ell-1} \frac{W}{k} \le W \ln \frac{\ell}{dn}.
    \end{equation*}
    Applying Markov's inequality yields
    \begin{align*}
        \Pr[X < (1-\fractionofcapacityweightratio)W] &= 1 - \Pr[X \ge (1-\fractionofcapacityweightratio)W] \\ 
        &\ge 1 - \frac{\E[X]}{(1-\fractionofcapacityweightratio)W} \ge 1 - \spaceleftmultiplier \ln \frac{\ell}{dn}.\qedhere 
    \end{align*}
\end{proof}
In the next Lemma, we use Lemma~\ref{lemma:small-singleround} to lower bound the total probability that a fixed fraction of a specific item will be packed.
\begin{lemma}[\cite{Albers2021}]\label{lemma:small-allrounds}
    Let $i \in \opt$ and $\variable_i$ be the fraction of item $i$ that is packed by $\algknapsack$. It holds that
    \begin{equation*}
        \Pr\left[\variable_i \ge \min\left\{\frac{\fractionofcapacityweightratio W}{s_i},\, \solfk_i\right\}\right] \ge 1-d + \spaceleftmultiplier \left[ 1-d - \left(1 + \frac 1 n \right)\cdot \ln \frac{1}{d} \right].
    \end{equation*}
\end{lemma}
\begin{proof}
    Summing up the probabilities from Lemma~\ref{lemma:small-singleround} over all rounds $\ell \ge dn + 1$ gives
    \begin{align*}
        \Pr\left[\variable_i \ge \min\left\{\frac{\fractionofcapacityweightratio W}{s_i}, 1\right\} \right] &= \sum_{\ell=dn+1}^n \Pr\left[\variable_i(\ell) \ge \min\left\{\frac{\fractionofcapacityweightratio W}{s_i}, \solfk_i\right\}\right]\\ &\ge \sum_{\ell=dn+1}^n \frac 1 n \left(1 - \spaceleftmultiplier\ln \frac{\ell}{dn}\right)\\
        &= \frac 1 n \left(n - dn - \spaceleftmultiplier\sum_{\ell=dn+1}^n \ln \frac{\ell}{dn} \right)\\
        &= 1 - d - \frac{1}{(1-\fractionofcapacityweightratio)n}\sum_{\ell=dn+1}^n \ln \frac{\ell}{dn} \\
        &\ge 1 - d - \frac{1}{(1-\fractionofcapacityweightratio)n}\left(n\cdot\ln \frac 1 d - n + dn + \ln \frac 1 d\right).
    \end{align*}
    The last inequality follows from $\sum_{\ell=dn+1}^n \ln \frac{\ell}{dn} = \left(\sum_{\ell=dn}^{n-1} \ln \frac{\ell}{dn} \right) + \ln \frac 1 d$ and, by Fact~\ref{fact:integralbounds}, $\sum_{\ell=dn}^{n-1} \ln \frac{\ell}{dn} \le \int_{dn}^n \ln \frac{\ell}{dn} \dif{\ell}$, which evaluates to $n\cdot \left(d - 1 + \ln \frac 1 d \right)$. 
    The claim follows by rearranging terms. 
\end{proof}
    Observe that the above lower bound on the probability becomes negative for any $\delta \ge 2 + \ln(d)/(1-d)$.
    Therefore, it is only safe to use it for those items $i$ that have a \textit{utilization}
    \[
        \utilization_i := \frac{s_i \solfk_i}{W} < \bar{\mu} := 2 + \frac{\ln d}{1-d},
    \] 
    where the utilization of an item measures its space consumption in the optimal fractional solution w.r.t.\ the total capacity and $\bar{\mu}$ denotes the maximum utilization allowed by our lower bound. In the following Lemma, we compute a lower bound on the expected profit obtained packing item $i \in \opt$ using $\algknapsack$, crucially relying on the ability of packing items fractionally. 
\begin{lemma}\label{lemma:small-packing}
    Let $i \in \opt$ and $\E_{i}[\algknapsack] = q_i \cdot  v_i \solfk_i$ be the expected profit obtained by $\algknapsack$ from the fraction of item $i$ that is packed in the optimal (offline) fractional solution. We have $q_i \ge q(\utilization_i) - o(1)$ where
    \begin{equation*}
         q(\utilization_i) = \frac{1}{\utilization_i}
            \left((1 - d)\cdot\min\{\utilization_i, \bar{\utilization}\} - \left(1-d-\ln \frac{1}{d}\right)\ln \left(1 - \min\{\utilization_i, \bar{\utilization}\}\right)\right).  
    \end{equation*}
\end{lemma}
    \begin{proof}
        Our goal is to compute the expected profit obtained from item $i\in \opt$ by \textit{summing up} the lower bounds on the probability that item $i$ is packed for a fraction $\variable_i \in (0, \solfk_i]$. To this end, let $N$ be an arbitrarily large parameter. We define $\delta_j = 1 - j/N$ for all $j \in \intInterCurly{0}{\pars}$. It follows that $\delta_j > \delta_{j-1}$. We may choose $N$ appropriately such that there is an index $k \in \intInterCurly{0}{\pars-1}$ with $\delta_k = \mu_i$. Note that $\mu_i$ is a rational number since the input data is rational.  
        Let us define $m_j=\frac{\delta_j W}{s_i}$ for $j\in \intInterCurly{k}{\pars}$. Recall that we consider a discrete probability space over the permutations of $[n]$. We consider the discrete random variable $X_i$ for the fraction of item $i$ that is selected by $\algknapsack$. Let $\Omega_i$ denote the finite set of values that $X_i$ can attain. We have
        \begin{align*}
            \E_{i}[\algknapsack] &= v_i \cdot \sum_{ x_i \in \Omega_i} x_i \cdot \Pr[X_i = x_i ] \\
            &\ge v_i \cdot m_k \cdot \Pr\left[X_i \ge m_k\right] + v_i \cdot \sum_{j=k+1}^{\pars} m_j \cdot \Pr\left[m_j \le X_i < m_{j-1}\right].
        \end{align*}
        Observe that $s_i \solfk_i \le \delta_k W$. Substituting $m_j = \frac{\delta_j W}{s_i \solfk_i} \cdot \solfk_i = \frac{\delta_j}{\delta_k}\cdot \solfk_i$ and using the fact that \[ \Pr\left[m_j \le X_i < m_{j-1}\right] = \Pr\left[X_i \ge m_j \right] - \Pr\left[X_i \ge m_{j-1} \right], \] yields
        \begin{align*}
              \E_{i}[\algknapsack] \ge v_i \solfk_i \cdot \Pr[X_i \ge m_k] +  v_i \solfk_i \cdot\sum_{j=k+1}^{\pars} \frac{\delta_j}{\delta_k} \left(\Pr[X_i \ge m_j] - \Pr[X_i \ge m_{j-1}]\right). 
        \end{align*}
        We rearrange the sum as follows
        \begin{align*}
             \E_{i}[\algknapsack] \ge v_i \solfk_i \cdot \left( \sum_{j=k}^{\pars-1} \frac{\delta_j - \delta_{j+1}}{\delta_k} \cdot \Pr[X_i \ge m_j ] + \frac{\delta_{\pars}}{\delta_k} \cdot \Pr[X_i \ge m_{\pars}]\right).
        \end{align*}
        By definition, we have $\delta_j - \delta_{j+1} = 1/\pars$, $\delta_{\pars} = 0$, and $\delta_k = \mu_i$.
        This implies that 
        \[
        q_i \ge \frac{1}{\mu_i N} \sum_{j=k}^{\pars-1} \Pr[X_i \ge m_j] \ge  \frac{1}{\mu_i N} \sum_{j=k'}^{\pars-1} \Pr[X_i \ge m_j].
        \]
        for any integer $k' \ge k$, which we will choose as the smallest index such that the lower bounds from Lemma~\ref{lemma:small-allrounds} are positive, i.e., $k' := \max\left\{k,\left\lceil N \cdot \frac{d - 1 - \ln d}{1-d} \right\rceil \right\}$. That is,
        \begin{align*}
              q_i & \ge \frac{1}{\mu_i \pars} \sum_{j=k'}^{\pars-1} \left[ (1-d) + \left(1-d-\ln \frac{1}{d} - \frac 1 n \ln \frac 1 d\right)\cdot\frac{1}{1-\fractionofcapacityweightratio_j} \right] \\
              &= \frac{1}{\mu_i} \left[ (1-d)\left(1 - \frac{k'}{\pars}\right) + \left(1-d-\ln \frac{1}{d} - \frac 1 n \ln \frac 1 d\right)\sum_{j=k'}^{\pars-1}\frac{1}{j} \right].
        \end{align*}
        It is easy to check that $\left(1-d-\ln \frac{1}{d} - \frac 1 n \ln \frac 1 d\right) \le 0$ for $d \in (0,1]$. Thus, using Fact~\ref{fact:integralbounds}, we deduce
        \begin{align*}
              \sum_{j=k'}^{\pars-1} \frac{1}{j} \le \sum_{j=k'}^{\pars} \frac{1}{j} \le \int_{k'-1}^{\pars}\frac{1}{t}\dif{t} = \ln \left(\frac{\pars}{k'-1}\right) = \ln \left(\frac{1}{1 - \left(1 - \frac{k'}{N}\right) - \frac{1}{\pars}}\right). 
        \end{align*}
        By plugging in the above upper bound and observing that $\delta_{k'} = 1 - k'/N$, we obtain
        \begin{align*}
              q_i \ge \frac{1}{\mu_i} \left[(1-d)\delta_{k'} + \left(1-d-\ln \frac{1}{d} - \frac 1 n \ln \frac 1 d\right)\ln \left(\frac{1}{1 - \delta_{k'} - \frac{1}{\pars}}\right) \right].
        \end{align*}
        Note that $\frac 1 n \ln \frac 1 d = o(1)$ and that we can achieve $\ln \left(\frac{1}{1 - \delta_{k'}}\right) \le \ln \left(\frac{1}{1 - \delta_{k'} - \frac{1}{\pars}}\right) \le \ln \left(\frac{1}{1 - \delta_{k'}}\right) + \varepsilon$ for any $\varepsilon > 0$, which  yields the desired lower bound for $\mu_i = \delta_k < \bar{\mu}$. Moreover, if $k' = \left\lceil N \cdot \frac{d - 1 - \ln d}{1-d} \right\rceil$, we can obtain $\bar{\mu} - \varepsilon \le \delta_{k'} \le \bar{\mu}$ for any $\varepsilon > 0$, which completes the proof.
    \end{proof}
This establishes the profit obtained by items in $\opt$ using $\algknapsack$. However, algorithm $\mathcal{A}$ can only benefit from $\algknapsack$ if algorithm $\algsecretary$ has not filled the knapsack completely. As we do not have any control over the (expected) size of the items packed by $\algsecretary$, we now condition on the event that it starts with an empty knapsack:
\begin{lemma}[\cite{Albers2021}]\label{lemma:empty}
    With a probability of at least $c/d$, no item is packed by $\algsecretary$.
\end{lemma}
Let $\xi$ denote the event of an empty knapsack after round $dn$. The following Lemma bounds the overall expected profit from $\algknapsack$'s packing for algorithm $\mathcal{A}$ by applying Lemma~\ref{lemma:empty}.
\begin{lemma}\label{lemma:small-final}
    We have $$\E_{\algknapsack}[\mathcal{A}] \ge \frac c d \sum_{i \in \opt} q_i \cdot v_i \variable_i. $$
\end{lemma}
\begin{proof}
    By Lemma~\ref{lemma:empty}, the probability of an empty knapsack after round $dn$ is at least $c/d$. Thus, we obtain 
    \begin{align*}
        \E_{\algknapsack}[\mathcal{A}] \ge \Pr[\xi]\cdot \E[\algknapsack | \xi] \ge \frac c d \sum_{i \in \opt} \E_i[\algknapsack] = \frac c d \sum_{i \in \opt} q_i \cdot v_i \variable_i, 
    \end{align*}
    by linearity of expectation and Lemma~\ref{lemma:small-packing}.
\end{proof}

\section{Competitive Ratio Analysis}\label{sec:cra}
We provide a unified analysis of the competitive ratio achieved by algorithm $\mathcal{A}$ which combines $\algsecretary$ and $\algknapsack$. 
We determine the best choice for the parameters $c$ and $d$ w.r.t.\ the bounds that we have proven before. Note that our bounds only make sense for $0 < c \le d \le 1$ and $\bar\mu > 0$. The latter yields an additional lower bound on $d$, which is the smaller of the two roots of $2-2d+\ln d$, say, $d_{\min} \approx \dminshort$.  In the following, we only consider the expected contribution by items that are contained in $\opt$. That is,
\[
\E_{\opt}[\mathcal{A}] \ge \sum_{i \in \opt} p_i \cdot v_i \solfk_i + \frac{c}{d} \sum_{i\in \opt} q_i \cdot v_i \solfk_i = \sum_{i \in \opt} \left(p_i+\frac{c}{d}q_i\right) \cdot v_i \solfk_i
.\]
Our strategy is as follows. We first reason about adversarial instances. We use these to determine the parameters $c$ and $d$ that yield the best competitive ratio that is permitted by the analysis of $\algsecretary$ and $\algknapsack$ above. Afterwards, we formally prove that we indeed covered the worst-case, i.e., we achieve the postulated competitive ratio with the particular values for $c$ and $d$ on all instances.
We have to take care of the following two situations:
\begin{enumerate}[(i)]
    \item $\opt$ only contains a single item.
    \item $\opt$ contains many items with a small utilization.
\end{enumerate}

In the former case, we have that $\opt$ must contain the most profitable item, which the secretary algorithm packs with a probability of $p_1$. Moreover, we start $\algknapsack$ with an empty knapsack with probability of at least $c/d$ and pack the most profitable item there with a probability of $q_1 \ge q(1)$. Hence, we obtain
\[
\E_{\opt}[\mathcal{A}] \ge \left( p_1 + \frac{c}{d} q_1 \right) v_1 \solfk_1 = \left( p_1 + \frac{c}{d} q_1 \right) \opt \ge \left( p(1) + \frac{c}{d} q(1) \right) \cdot \opt,
\]
where we now and in the remainder of this section omit the $o(1)$ for better readability. Let $1/z$ denote the competitive ratio that we want to show. By the consideration above, we have that
\begin{equation}\label{eq:singleitem}
z \le p(1) + \frac{c}{d} q(1).
\end{equation}
For the second case, algorithm $\algknapsack$ gives a lower bound of $c/d \cdot q(\mu)$ on the fraction  that we pack of each item, w.r.t. its utilization in the optimal fractional solution, regardless of the cardinality of $\opt$. Since $q(\mu)$ is decreasing for $\mu \in (0,1)$ and any choice of $d \in (d_{\min},1)$, we have the following constraint for the competitive ratio
\begin{equation}\label{eq:manyitems}
z \le \frac{c}{d} \cdot q(0).
\end{equation}
In fact, the optimum of the resulting optimization problem 
\begin{maxi*}
	  {c,d,z}{z}{}{}
	  \addConstraint{z}{\le c\cdot\ln\frac{d}{c} + \frac{c}{d}\cdot\left( 2-2d + \ln d - \left(1-d-\ln \frac 1 d \right)\cdot \ln \left( \frac{\ln\frac{1}{d}}{1-d} - 1 \right)\right)}{}
	  \addConstraint{z}{\le \frac{c}{d} \cdot \left( 2-2d+\ln d\right)}{}
	  \addConstraint{0}{< c \le d \le 1}{}
\end{maxi*}
is attained when the two upper bounds on $z$ are equal, which yields $c$ in dependence on $d$. Hence, $z$ is determined by a univariate function in $d$. It has a local maximum in the interval from $d_{\min}$ to $1$, which is also a global maximum in that interval. A numerical computation yields that the maximum is attained for some $d\approx \dtreshfull$. This yields $c\approx \ctreshfull$ and a competitive ratio of $\fullratio$. 
    
In the following, we will show that the competitive ratio for all other cases is not worse for the parameters above completing the proof of Theorem~\ref{thm:main}.
\begin{lemma}
    For all $\card \in [n]$, the algorithm $\mathcal{A}$ is $\frac 1 z$-competitive for \[z=\frac c d \left(2 - 2d  - \log(1/d)\right) > \frac{1}{\ratio},\] $c=\ctresh$, and $d=\dtresh$.
\end{lemma}
\begin{proof}
    By Lemmas~\ref{lemma:largeprob}~and~\ref{lemma:small-final}, the profit obtained by items in $\opt$ is at least $\sum_{i\in \opt} \left( p(i) + \frac c d q(\utilization_i)\right) \cdot  v_i \solfk_i$. 
    Clearly, the case $\card = 1$ follows from the above optimization problem. For the general case $\card \ge 2$, the minimum coefficient over all terms in the summation, for the given $c$ and $d$, may be smaller than the desired competitive ratio. However, if their average is larger than $z$, then we can exploit this as follows. The idea is that we transport \textit{excess} from items with a smaller index (i.e., larger contribution to $\opt$), to items with larger index that may have a deficit. That is, we can redistribute the excess at item $1$ to any larger indexed item. Note also that for any $\mu_i$, it holds that $c/d \cdot q(\utilization_i) \le z$ and that we can assume without loss of generality that $\mu_1 \ge \ldots \ge \mu_\card$ as any other assignment of the $\mu_i$'s would give an higher expected value. Therefore, we deduce the following sufficient condition for our lemma
    \begin{align*}
        p(1) + \frac{c}{d}\sum_{i=1}^\card q(\utilization_{i}) \ge z \cdot \card.
    \end{align*}
    In particular, substituting $q(\utilization_i)$ with its definition in $\sum_{i=1}^\card q(\utilization_{i})$, we have
    \begin{align*}
                (1 - d)\cdot \min\left\{1,\frac{\bar{\utilization}}{\utilization_1}\right\} &+ \left(1-d-\ln \frac{1}{d}\right)\ln \left(\frac{1}{1 - \min\{\utilization_1, \bar{\utilization}\}}\right) \\
                &+ \sum_{i=2}^\card \left(1 - d + \frac{1}{\utilization_i}\left(1-d-\ln \frac{1}{d}\right)\ln \left(\frac{1}{1 - \utilization_i}\right)\right).
    \end{align*}
    Assuming that $\utilization_1 \le \bar{\utilization} \approx \mubar$, we can rewrite the above expression as follows
    \begin{align*}
        \card\cdot(1-d) - \left(1-d-\ln \frac{1}{d}\right)\sum_{i=1}^\card \frac{1}{\utilization_i} \ln \left(1 - \utilization_i \right).
    \end{align*}
    Recalling that $\left(1-d-\ln \frac{1}{d}\right) < 0$, for $d \in (0,1]$, and that $\sum_{i=1}^\card \utilization_i \le 1$, we observe that $\sum_{i=1}^\card \frac{1}{\utilization_i} \ln \left(1 - \utilization_i \right)$ is a separable concave function over the polyhedral domain $\{ \utilization \in [0, \bar{\utilization}]^\card \: : \: \mathds{1}^T \utilization \le 1 \}$, which is minimized at the boundary, i.e., choosing $\utilization_1 = \bar{\utilization}, \utilization_2 = 1 - \bar{\utilization}$ and $\mu_j \rightarrow 0$ for $j \ge 3$. Note that for $\utilization_j$ tending to zero, the term $\ln(1-\utilization_j)/\utilization_j$ converges to $-1$ from below. Meaning that $\sum_{i=1}^\card \frac{1}{\utilization} \ln \left(1 - \utilization \right) \ge \frac{1}{\bar{\utilization}} \ln \left(1 - \bar{\utilization} \right) + \frac{1}{1 - \bar{\utilization}} \ln \left(\bar{\utilization} \right) - (\card - 2)$, resulting in the following condition
    \begin{align*}
        p(1) + \frac{c}{d}\left(\card-\card d - \left(1-d-\ln \frac{1}{d}\right)\cdot \left(\frac{1}{\bar{\utilization}} \ln \left(1 - \bar{\utilization} \right) + \frac{1}{1 - \bar{\utilization}} \ln \left(\bar{\utilization} \right) - \card + 2 \right)\right)\\
        = \underbrace{p(1) - \frac{c}{d}\left(1-d-\ln \frac{1}{d}\right)\left(2 + \frac{1}{\bar{\utilization}} \ln \left(1 - \bar{\utilization} \right) + \frac{1}{1 - \bar{\utilization}} \ln \left(\bar{\utilization} \right) \right)}_{\approx \firstcondition} + z\card,
    \end{align*}
    after plugging in the definition of $z$. Let us now consider the case in which $\utilization_1 > \bar{\utilization}$. Then, the term $\sum_{i=1}^\card q(\utilization_{i})$ can be rewritten as follows
    \begin{align*}
        (1 - d)\cdot \frac{\bar{\utilization}}{\utilization_1} &+ \frac{1}{\bar{\utilization}}\left(1-d-\ln \frac{1}{d}\right)\ln \left(\frac{1}{1 -  \bar{\utilization}}\right)\\ &+ (\card-1)(1-d) - \left(1-d-\ln \frac{1}{d}\right)\sum_{i=2}^\card \frac{1}{\utilization_i} \ln \left(1 - \utilization_i \right),
    \end{align*}
    where the sum is minimized for $\utilization_2 = 1 - \utilization_1$ and $\mu_j \rightarrow 0$ for $j > 2$. The resulting function in $\utilization_1$ is decreasing over $(\bar{\utilization},1]$ for our choice of parameters. Thus, it is lower bounded by
    \begin{align*}
        &\begin{aligned}
            p(1) &+ \frac{c}{d}\left((1 - d)\bar{\utilization}+ \left(1-d-\ln \frac{1}{d}\right)\ln \left(\frac{1}{1 -  \bar{\utilization}}\right)\right)\\ 
             &+ \frac c d (\card-1)\left(2-2d-\ln \frac{1}{d}\right)
        \end{aligned}\\
        &\begin{aligned}
         = \underbrace{p(1) + \frac{c}{d}\left((1 - d)\bar{\utilization}+ \left(1-d-\ln \frac{1}{d}\right)\ln \left(\frac{1}{1 -  \bar{\utilization}}\right)
         -2+2d+\ln \frac{1}{d}
         \right)}_{\approx \secondcondition} + z\card.
        \end{aligned}
    \end{align*}
    Both conditions are met for our choice of $c,d$, concluding the proof.
\end{proof}

\section{Future Directions}
Potential improvements and future directions are:
\begin{enumerate}[(i)]
\item Better bounds on the probability of picking the $i$-th most valuable item in the secretary algorithm. We use the corresponding lower bounds from~\cite{Albers2021}; however, the first one already suffices for the single-item case, and an improvement of the many-item case would require that the amortized probability of picking the $k$-th most-valuable item does not go to $0$ as $k$ gets large.
\item Close the probability gap in the analysis. With our choice of $c$ and $d$, we account for a probability of about 11\% that the secretary algorithm packs the most profitable item, and for a probability of about 79\% that the third phase starts with an empty knapsack, which we use to condition the expected profit of the knapsack algorithm -- leaving a gap of about 10\% that is lost for our analysis.
\item Interleaving of secretary and knapsack algorithm. Instead of running two separate phases for completing the secretary and the knapsack algorithm, they could run in parallel and the decisions on the current item could be based on a combination of both opinions (e.g., min, max, or coin flip).
\item Consequences for the online knapsack problem. Our focus was to improve the competitive ratio for the online fractional knapsack problem. Additionally, it would be interesting to investigate whether our result leads to new insights about the integer problem, e.g., w.r.t.\ the analysis of randomized rounding or in other closely related settings.
\end{enumerate}

\bibliographystyle{splncs04}
\bibliography{references}
\end{document}